# Construction of a Small-Scale Vacuum Generation System and Using It as an Educational Device to Demonstrate Features of the Vacuum

Osama A. Marzouk[1], Walid Ali Marjan Haje Rhaim Jul[1], Amjad Masoud Khalfan Al Jabri[1], Hamed Aamir Mohammed Aamir Al-ghaithi[1]

[1]College of Engineering, University of Buraimi, Sultanate of Oman

Correspondence: Osama A. Marzouk, College of Engineering, University of Buraimi, Post Code 512, Al Buraimi, Sultanate of Oman.

**Abstract**

We developed a vacuum generation system composed of a reciprocating compressor (3 tons of refrigeration) with an inverted-function that is ready to be hooked flexibly to a gas-tight container to create an evacuated enclosed atmosphere, without strict limitation of the size of that container. The evacuated container (or vacuum chamber) can serve in different purposes such as educational demonstration of the vacuum properties, extraction of perfumes from herbal resources, and preserving food. We tested the device and found it can reach a vacuum level of 26 inches of mercury in an environment with an atmospheric pressure of 28.5 inches of mercury. We compared the performance of our vacuum device to a rotary-vane vacuum pump of ¼ horsepowers and found that the vacuum pump reaches a set test vacuum level of 25 inches of mercury before the compressor. We then demonstrated experimentally some features of the vacuum using the inverted compressor or the vane vacuum pump. These experiments serve some topics in physics for school students as well as two core subjects of mechanical engineering, namely fluid mechanics and thermodynamics.

## 1. Introduction

The work presented here is based on a graduation project in Mechanical Engineering, where the involved team aimed at designing a vacuum chamber and connecting it to an inverted-function compressor. Then, the resulting vacuum generator set is operated as an educational device to demonstrate practically and effectively some concepts to school students. The experiments used with the vacuum generator are water vaporization under vacuum, isothermal volume increase (expansion) due to vacuum, and wave transmission in vacuum.

One of the main ideas is to convert a reciprocating compressor into a vacuum machine, which sucks the air from one side and blows it to the other side. The suction side is attached to a sealed chamber to prevent any air leakage as this can lead to a loss of vacuum inside the chamber. The final device constructed is shown in Figure 1.

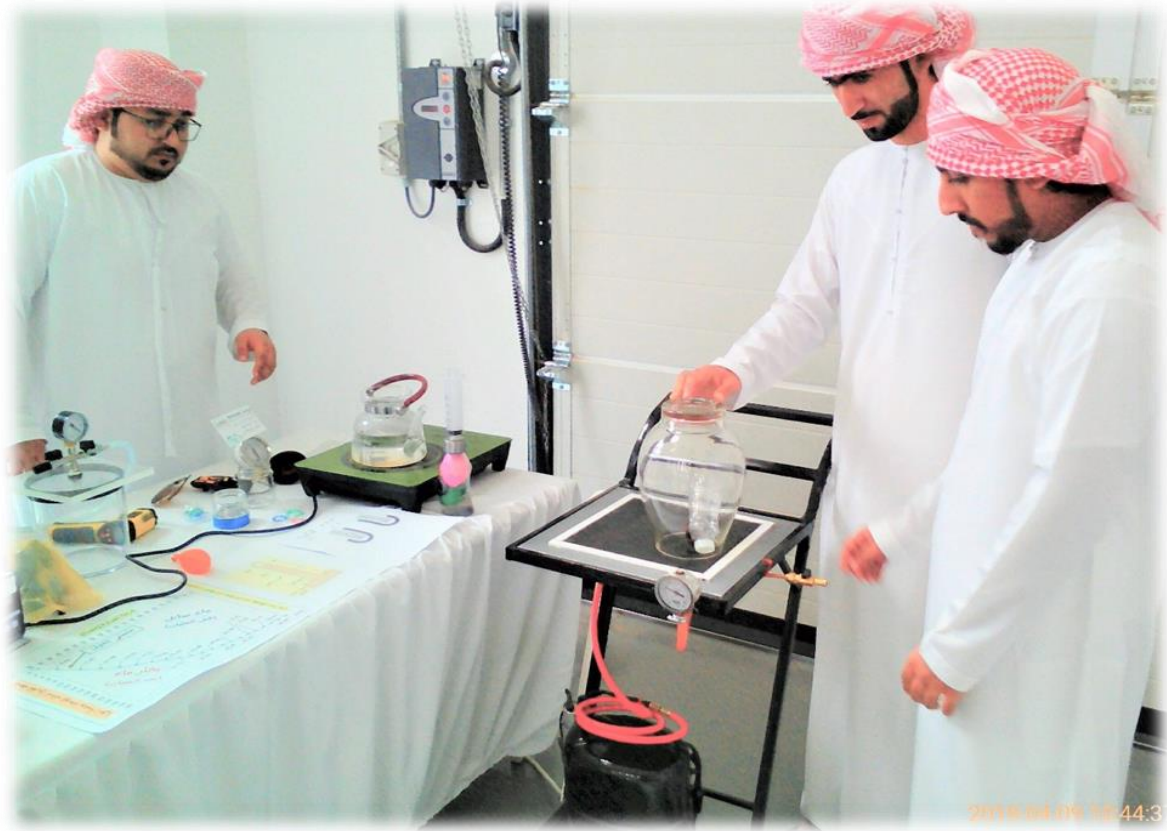

Figure 1. Constructed vacuum generator with a movable cart and team of the work (Mr. Walid to the left, M. Amjad to the right, and Mr. Hamed in the middle)

The vacuum generator is fitted with a carrying cart for easy mobility; it provides a supporting frame. The top of the cart has a flat gasket with a protruding tip of a copper tube connected to a cross fitting below the gasket. The other terminals of the cross are: one to the inverted compressor (gas suction), one to a vacuum gauge, and one to a shut-off valve for vacuum relief. A chamber can be placed while flipped upside down over the protruding suction tube and then one needs to press gently for achieving an initial seal and start operating the inverted compressor to start evacuating the sealed chamber. During evacuation, the seal increases as the chamber rim is stuck to the gasket. The relief valve should be closed during the evacuation process. The gasket was made from the tube of a car tire. The device operated successfully and we avoided the need to purchase a vacuum pump that can be costly and not readily available locally. However, an air conditioning compressor is a common component and one can get a used one with a much less price than the one for a new vacuum pump, even a small-size unbranded one.

We explored some other similar works related to small-scale vacuum generation, and found that different methods were used to create vacuum and different designs exist for the vacuum chambers. There are several applications of the generated vacuum (Jayaram and Gonzalez, 2011; Tripathi et al., 2012; Fujimoto et al., 2017), and a specific application may favor one method over the others.

We refer the reader to a useful online article about how to make a vacuum chamber by Mr. Brean (posted online at http://equation.com), who wrote about how to make a vacuum chamber out of simple components that can be found easily. He stared with an explanation for a vacuum chamber, being a closed space where normal pressure does not exist, but a much lower pressure level is reached. His vacuum chamber can be used to create certain components of computers. For us, our main goal is to explore the properties of vacuum. The materials that Mr. Brean suggested to use or actually used are:

1) Mason jar

2) Metal-cutting tool

3) Super glue

4) Rubber stopper

5) Vacuum pump

The procedure of Mr. Brean is to cut the top of the jar so it just fits a rubber stopper and then insert the rubber stopper into the top of the jar then place the items to be exposed to vacuum inside the jar and screw the top tightly and then connect the nozzle of the vacuum pump to the jar through the stopper and turn on the vacuum pump to remove the air from the jar.

This idea is a simple way to create a vacuum chamber, but using a vacuum pump was not favored by us. So, we started looking for a replacement for the vacuum pump.

Another practical work we found belongs to Mr. Ramesh, who provides air conditioning repair and installation services in India. He is running a company with the name BENT AIR COOL in Mumbai. He has an interesting video on YouTube (www.youtube.com), where he describes how one can use an old air conditioning compressor to evacuate an air conditioning system. Mr. Ramesh used an air conditioning compressor with 1/5 tons of refrigeration (about 0.94 horsepowers) and connected it to a vacuum gauge and described how one can use the suction line of the compressor to suck the left moisture in the air conditioning tubes and then he described how one can connect the discharge line to a tube in order to examine if it is leaking (compressor lubricant) or not.

Inspired by the above example, we also decided to replace a vacuum pump by a reciprocating air conditioning compressor, which can work as a vacuum pump upon inverting its ports connectivity.

**2. Theoretical Overview**

We give an overview about a common type of vacuum pumps and a common type of air conditioning and refrigeration compressors, which are involved in our work. We cover here a category of the positive displacement vacuum pumps, which are the main pumps used in vacuum systems (Jousten et al., 2016). Their working principle is opening a small cavity and allowing the gas to flow into it from the chamber being evacuated. Then, the vacuum pump seals and exhausts the gas out to the atmosphere. The level of vacuum depends on the mechanical motion. Biologically, one finds one of these pumps in her/his own body, which is the diaphragm muscle, the main breathing muscle (Bordoni et al., 2016), that expands and contracts the chest cavity, resulting in an increase and decrease of the volume of the lungs, respectively.

Positive displacement vacuum pumps are classified into some categories, based on the motion mechanism. The category of interest to us is called rotary-vane pump (Surhone et al., 2010). In this category, a rotor revolves at a high-speed inside a cylindrical case, and it is set slightly off-centered in a shaft. The rotor has vanes that touch the case wall at one side during rotation, and they suck the gas from the case (and the chamber connected to it from outside). Because one side of the rotor is off the center, there is a pressure difference between the two sides, and the rotor blows the gas to the outside atmosphere through a discharge port connected to the higher-pressure side of the case.

A reciprocating compressor belongs to the positive displacement machinery, but it operates to pressurize a system rather than to evacuate it. This category of compressors has a piston moving upward and downward inside a cylinder (loch and Hoefner, 1996). The downward motion reduces the pressure in the cylinder zone located above it, thereby generating a vacuum in that zone. Due to the pressure difference between the low vacuum inside and the atmospheric pressure (or higher pressure level in general) outside, a suction valve opens and brings the air (or gas in general) in and as the piston moves up, it pushes the gas out of the cylinder through a discharge valve.

In air conditioning applications, a compressor is not always specified by its own but it can be specified by the air conditioning system it fits, using a rating value called tons of refrigeration (TR), or simply tons. This rating is related to the rate of cooling and removal of heat. One ton of refrigeration corresponds to an air conditioning unit capable of removing 12,000 BTU of heat in one hour (Bevan, 2013), where BTU is the British thermal unit of heat. Likewise, a rating of 2 tons of refrigeration means a cooling rate of 24,000 BTU/h. One ton of refrigeration as a power unit is equivalent to 3.52 kilowatts--kW (Dincer, 2018) or 4.72 horsepowers--hp. We can relate the removal of heat to the size of a compressor (the volume of its cylinder). The larger the cylinder, the higher the refrigeration tonnage, and the faster the vacuum is obtained.

**3. Constructed Vacuum System**

We utilized an air conditioning reciprocating compressor with 3 tons of refrigeration rating, and it is shown in Figures 2 and 3. We bought this compressor as a used unit (not a new one) from a local shop, and thus we do not have its full specifications. However, based on the markings on its body, we know it was made in the USA. By connecting the compressor's discharge port to the atmosphere and the suction port to the sealed chamber, the compressor function is inverted and it acts as a vacuum machine.

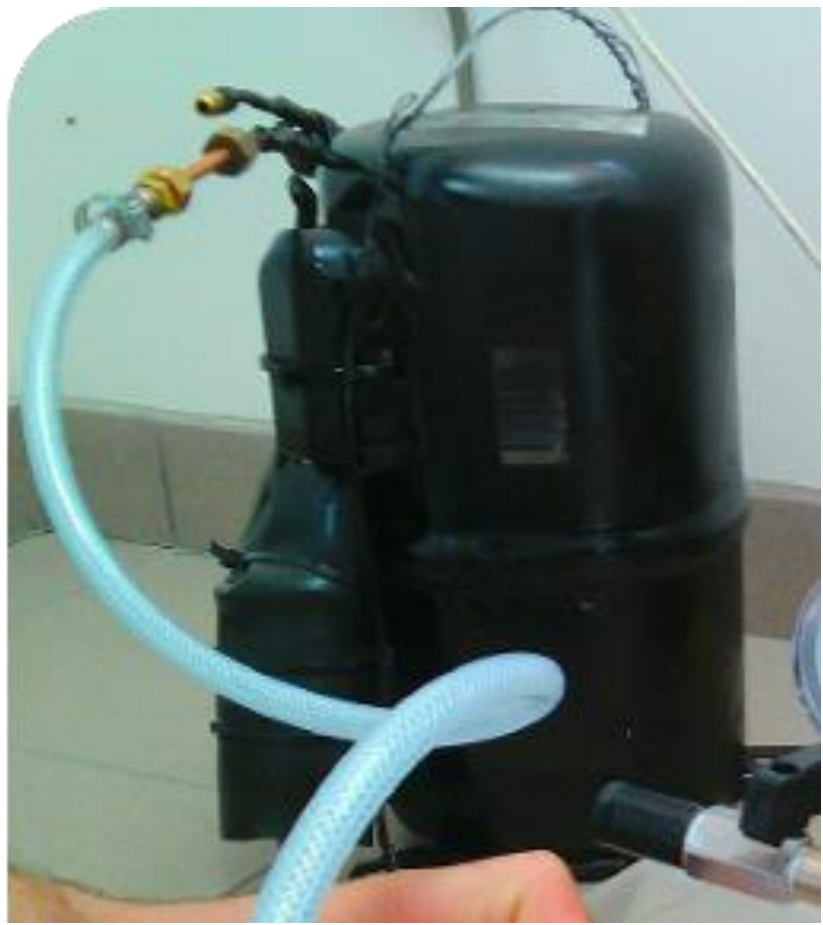

Figure 2. The compressor used in our work, before assembling it within the vacuum generator

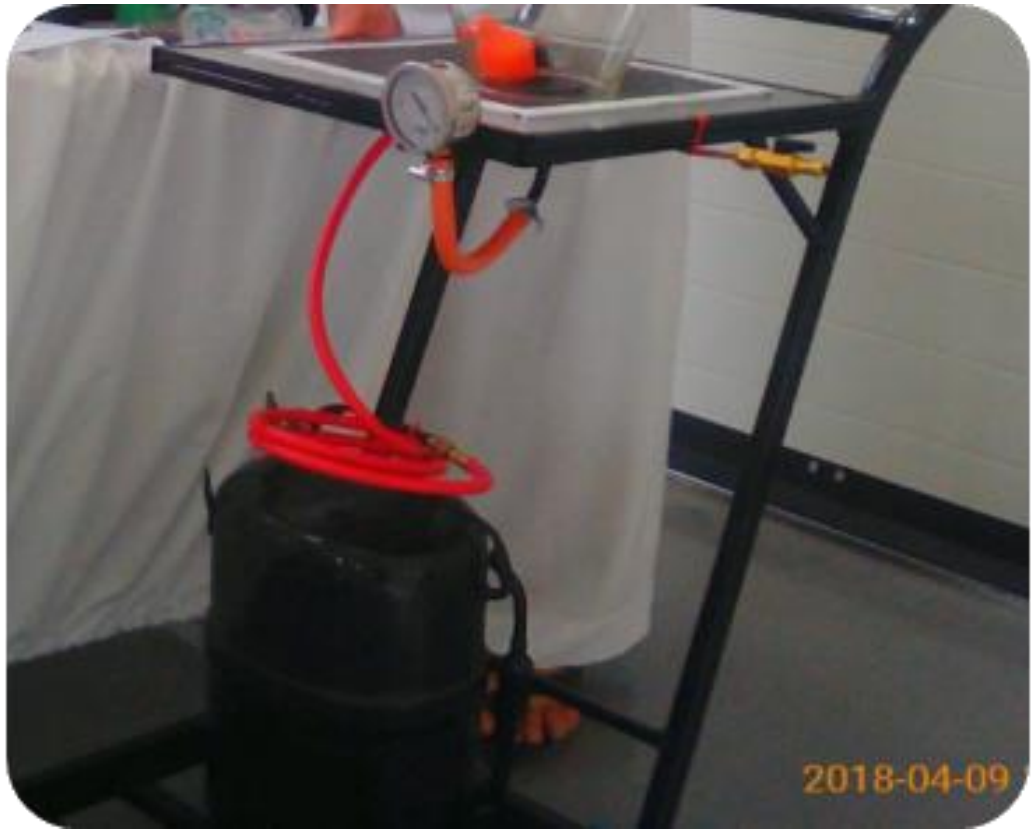

Figure 3. The compressor used in our work, after assembling it within the vacuum generator

For the vacuum chamber, we decided to use ready glass containers (large vases) in the market. To use the chamber, it should be flipped upside down such that its rim rests on a horizontal gasket. The rim surrounds a suction port connected to the inverted compressor. This achieves sealing that increases as evacuation progresses. It also allows flexibility with installing chambers since any gas-tight container with an opening that can rest on the gasket can be used as a chamber without any customization. The gasket has a square shape with a side length of 26 cm. It was made from the tube of a car tire.

In addition to the compressor, the items used to construct the vacuum system include:

1) Vacuum gauge (Winters PFQ series, glycerin filled, accuracy ±1.5% of full scale)

2) Wood plate (square: 40cm ×40cm) with a hole of ½ inches to pass a suction copper tube

3) Gasket (square: 26cm ×26cm) to provide sealing for the chamber

4) Copper pipes and fittings (including a cross for: compressor, chamber, gauge, and relief)

5) All-purpose glue and silicone (for assembly work)

6) Electric on-off switch (to allow operating the compressor)

7) Discharge shut-off valve (to open the chamber to the atmosphere and relief the vacuum)

8) Clear hose connected to the discharge of compressor (to store/collect any oil leak)

9) Electric wires (for wiring the compressor and making a power chord for it - 230 V AC)

10) Gas cylinder hose (as a flexible connector between the gauge and the cross)

Considering the heavy weight of our compressor, we decided to make a trolley that can carry it along with the entire system. We designed a two-floor, four-wheel trolley that carry the chamber at the top floor and connect all the pipes and wires to the inverted compressor, which is at the bottom floor.

**4. Comparison With a Vane Vacuum Pump**

We compared between our inverted compressor and a rotary-vane vacuum pump as a benchmarking step. The vacuum pump is an unbranded Chinese unit, having a model code of RS-1. It is a single-stage pump with free-air flow of 2.5 cubic feet per minute (CFM) and ¼ hp of power. A photo for the pump is given in Figure 4.

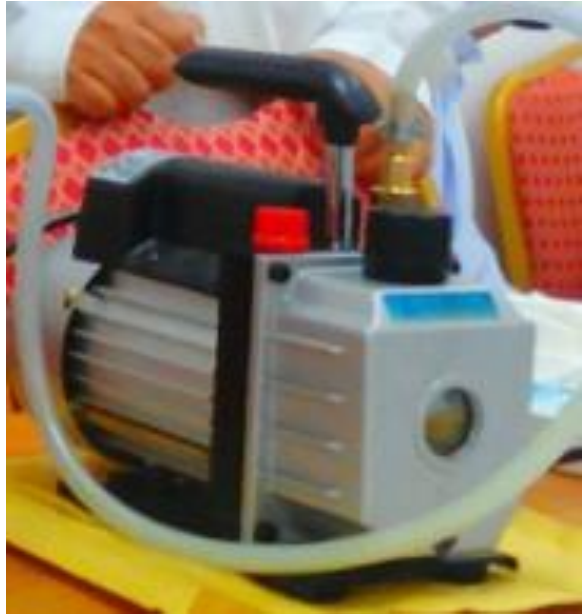

Figure 4. A benchmarking rotary-vane vacuum pump

The comparison focuses on assessing which vacuum machine reaches a vacuum of 25 inches of mercury (inHg) faster while evacuating a 2-liter cylindrical chamber. The benchmarking test was repeated three times and the results see consistency as given in Table 1 as times elapsed to reach the set vacuum level in seconds.

Table 1. Comparison between the inverted compressor and the vane vacuum pump

| Pump | Compressor |
|---|---|
| First attempt = 10.32 seconds | First attempt = 13.80 seconds |
| Second attempt = 10.46 seconds | Second attempt = 13.72 second s |
| Third attempt = 10.59 seconds | Third attempt = 13.40 seconds |
| Average = 31.77/3 = 10.45 seconds | Average = 40.92/3 = 13.64 seconds |

The vane vacuum pump reached the set level of vacuum in less time than the compressor, and on average it is 23% faster than the compressor, where (13.64s – 10.45s) / 13.64s = 0.234.

We add here that the maximum vacuum can be reached by both machines is about 25.6 inHg for the inverted compressor and 26.1 inHg for the rotary-vane vacuum pump. This corresponds to about 2 inHg (inverted compressor) or 2.5 inHg (vane vacuum pump) of absolute pressure, given that the local atmospheric pressure at the time of performing this work was about 28.6 inHg. This is less than the standard value of 29.9 inHg, but the local area here at the University of Buraimi is at an elevation of about 360 m above sea level, which causes a drop in the atmospheric pressure from the standard value.

**5. Performed Educational Experiments**

We designed and implemented three experiments that can be used to teach school students at the appropriate grade about vacuum and its features. These experiments are:

*5.1 Experiment 1: Cold Boiling*

Cold boiling, vacuum boiling, or cavitation is a phenomenon that is not encountered frequently in our normal life. Vacuum boiling refers to the change of water phase from liquid to vapor due to depressurization, not due to heating as in common boiling conditions. While water boils at 100 °C under a normal pressure of 1 atm, which is equal to 101,325 Pa, if we decrease the pressure surrounding the water it will boil at a lower temperature. This means it is possible to boil very cold water near zero degrees Celsius if its pressure is near absolute vacuum. However, with the inverted compressor or vacuum pump that we have, we cannot reach that high level of vacuum. Therefore, the water needs to be heated to some extent to raise its temperature, which facilitate boiling at the level of vacuum we can have (about 2.5 inHg absolute vacuum).

We measured the local atmospheric pressure using a Greisinger GPB 3300 Digital Barometer (Made in Germany,

accuracy ±2 mbar), and found it to be about 97,000 Pa, which corresponds to 28.64 inHg. Given the relation between saturation pressure and saturation temperature (Cengel and Boles, 2015), we constructed a visual representation for the boiling locus in Figure 5. With our maximum vacuum level of about 26 inHg, we need to raise the temperature of water to at least 45 ℃ in order to have boiling. This is still remarkably below the standard boiling point of 100 ℃. We ran this experiment using the vane vacuum pump.

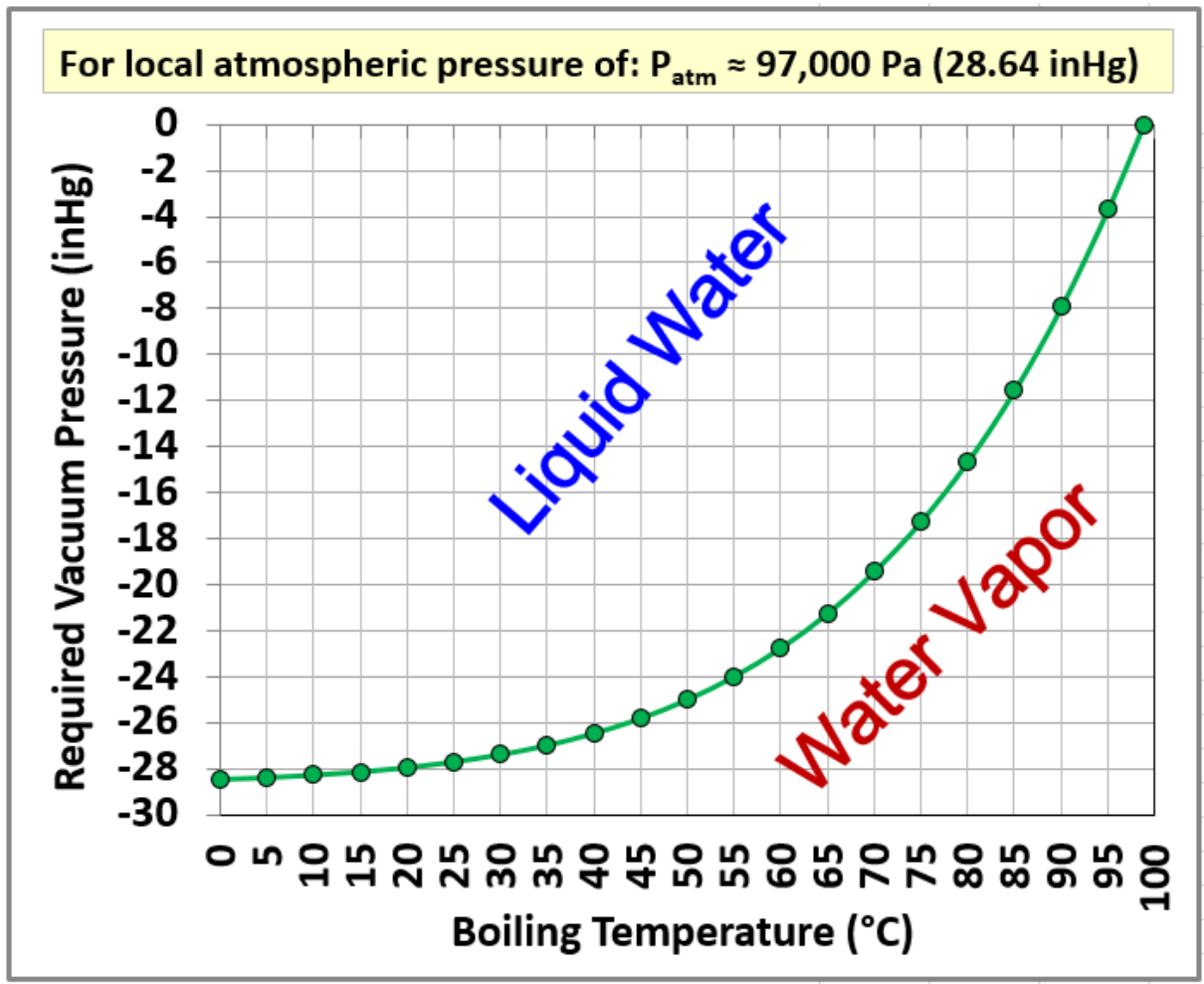

Figure 5. Vacuum pressure (represented as negative values) needed for boiling water as a function of its temperature

The items we needed to perform this experiment are:

1) Water in a cup

2) Water heater (we used an electric hot plate)

3) Thermometer (to monitor the water temperature)

Figure 6 is a photo for a 2-liter acrylic vacuum chamber (with a customized rim sealing ring provided by the seller: ZEEVAC LTD, in the UK) during the cavitation of warm water placed inside a small container within the chamber.

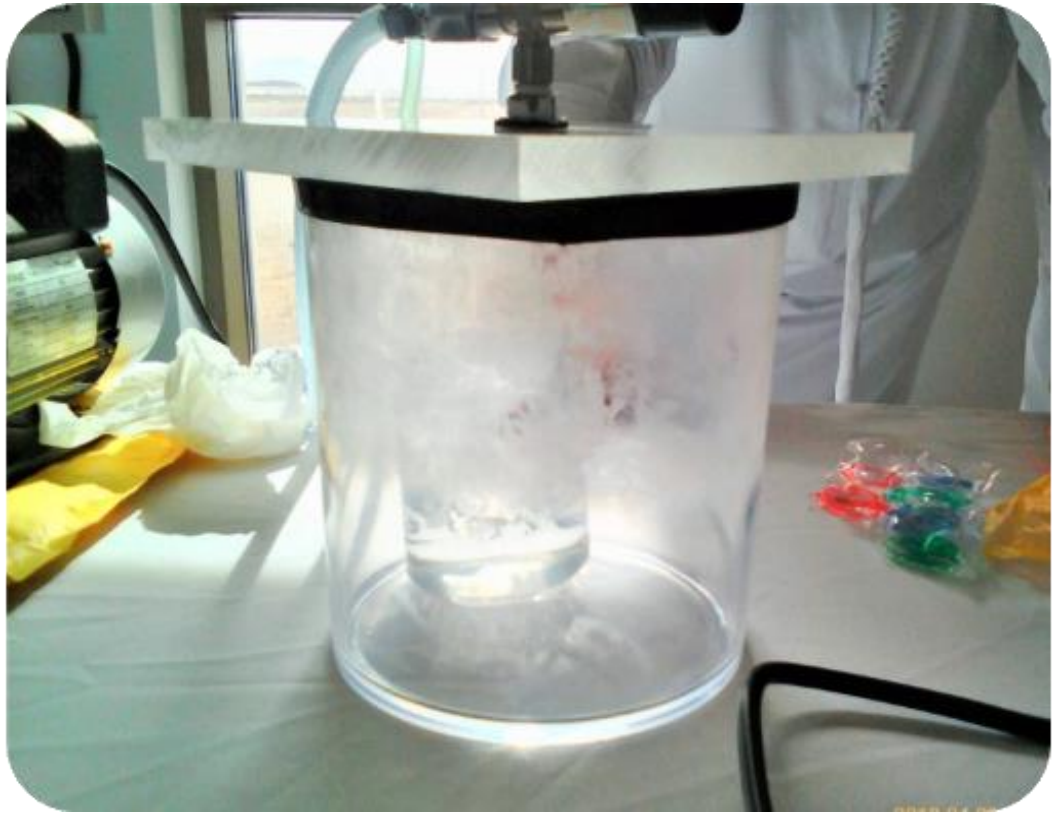

Figure 6. Cold boiling inside one of our chambers

*5.2 Experiment 2: Pressure-Volume Relationship*

The second experiment that we have chosen to perform is the volume increase (expansion) under vacuum. From Boyle's law, a classical ideal-gas law in thermodynamics (Vallance, 2017), pressure is inversely proportional to the volume at a constant temperature. The volume of a flexible closed system increases as vacuum starts. The system might burst if it expands excessively. We used a regular party balloon (about 12 inches for the size) and a squeezed plastic water bottle (size ½ liters). We ran this experiment using the inverted compressor and the constructed vacuum system.

The procedure to perform this experiment is:

1) Trap some air inside the balloon and tie it.

2) Put the balloon in the vacuum chamber and see its inflation under vacuum.

3) Replace the balloon with a squeezed water bottle and observe its enlargement under vacuum.

Figure 7 shows the experiment during the inflation of the balloon inside a glass vacuum chamber.

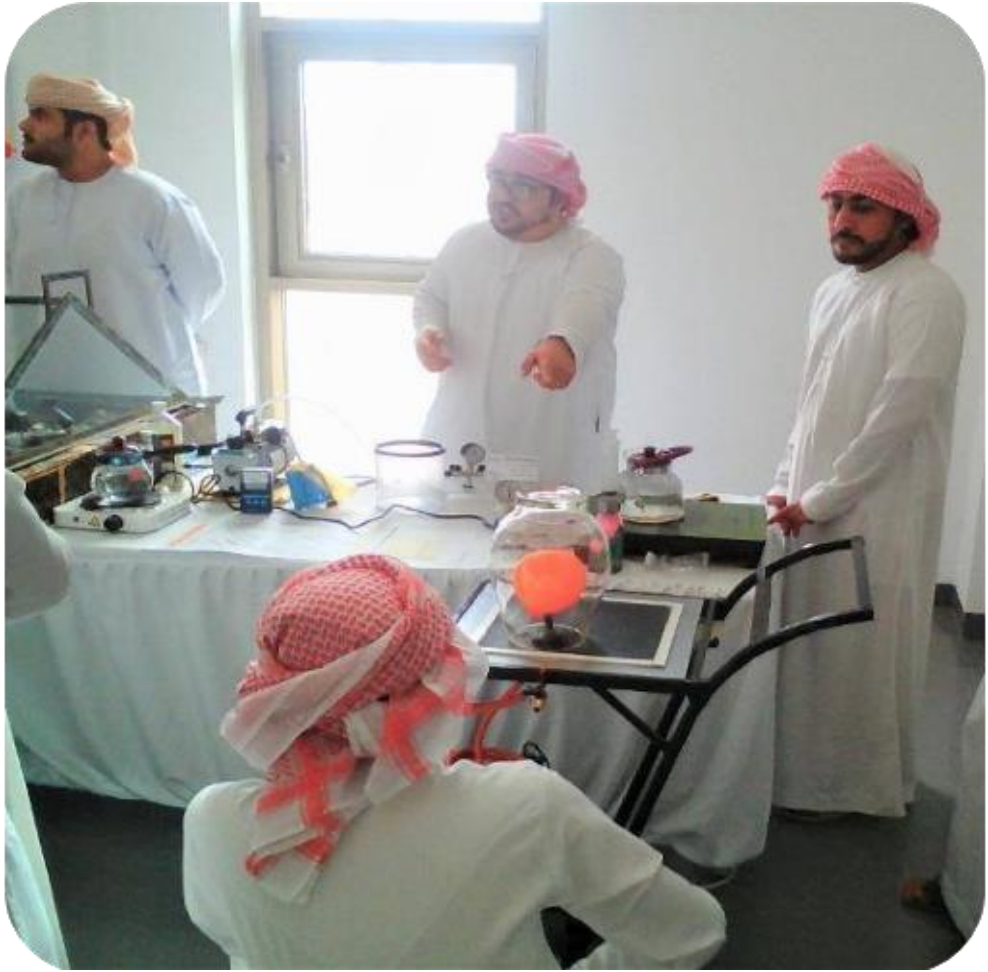

Figure 7. The balloon expansion experiment

*5.3 Experiment 3: Wave Propagation*

The third experiment that we performed is placing a sound-wave source (a bicycle electric horn in our case) and an electromagnetic-wave source (a mobile phone in our case) inside the vacuum chamber and observing if these waves can travel through the vacuum. We ran this experiment using the vane vacuum pump.

The procedure to perform this experiment is:

1) Put the sound source inside the vacuum chamber.

2) Turn on the vacuum pump and observe the sound level before and after the vacuum.

3) Put a mobile phone inside the chamber and evacuate it and see if it can receive a call.

Conforming to principles of physics, the sound level was weakened in the evacuated environment since sound waves are pressure waves that require a transmission medium, whereas the mobile phone could receive the call while in the evacuated chamber since electromagnetic waves do not require a medium and can travel through vacuum. Figure 8 is a photo for the horn sound source while placed in the vacuum chamber. The horn used has a LED source of light, which helps indicating it is operating and the sound weakness is not due to any malfunctioning of the horn itself.

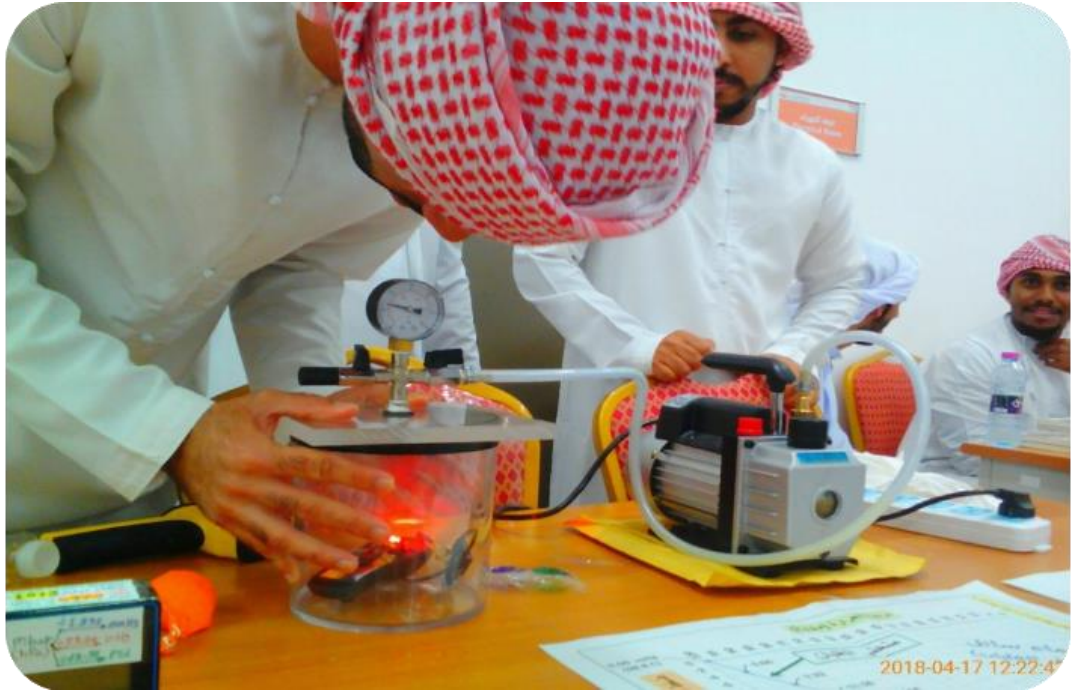

Figure 8. The sound wave propagation experiment

**6. Conclusions**

We built a mobile vacuum generation system by inverting the function (swapping the ports connections) of a large air conditioning compressor, and assembling it with a chamber-ready sealing gasket and a two-floor trolley. Vacuum chambers in that system are bottom-connected and thus need to be flipped vertically while in use. We also used another vacuum system but one purchased ready (we just assembled the components), that uses a small rotary-vane vacuum pump and a top-connected 2-liter acrylic vacuum chamber. We compared the performance of the vane vacuum pump and the inverted-function compressor, and used these systems to run three experiments that demonstrate some features of the vacuum.

The experiments that we demonstrated are water vaporization at low temperatures (cavitation), volume increase due to pressure decrease without temperature change (in compliance with Boyle's law), and the inability of sound wave to travel in vacuum in contrast to electromagnetic waves. More experiments can be developed, and either of the two vacuum systems can be used as a teaching facility in relevant schools (middle and/or high stage) as well as engineering students who study topics in fluid mechanics, thermodynamics, and plumbing. We have actually performed these experiments in front of high-school students and college students in the local area of Al Buraimi in Oman, and it was very pleasant to see some being excited and taking photos for the setup by their personal mobile phones.